# Protein single-model quality assessment by feature-based probability density functions

Renzhi Cao[1] and Jianlin Cheng[1,2,3,*]

[1]Department of Computer Science, University of Missouri, Columbia, MO 65211, USA
[2]Informatics Institute, University of Missouri, Columbia, MO 65211, USA
[3]Bond Life Science Center, University of Missouri, Columbia, MO 65211, USA

[*]chengji@missouri.edu

**ABSTRACT**

Protein quality assessment (QA) has played an important role in protein structure prediction. We developed a novel single-model quality assessment method - Qprob. Qprob calculates the absolute error for each protein feature value against the true quality scores (i.e. GDT-TS scores) of protein structural models, and uses them to estimate its probability density distribution for quality assessment. Qprob has been blindly tested on the 11th Critical Assessment of Techniques for Protein Structure Prediction (CASP11) as MULTICOM-NOVEL server. The official CASP result shows that Qprob ranks as one of the top single-model QA methods. In addition, Qprob makes contributions to our protein tertiary structure predictor MULTICOM, which is officially ranked 3rd out of 143 predictors. The good performance shows that Qprob is good at assessing the quality of models of hard targets. These results demonstrate that this new probability density distribution based method is effective for protein single-model quality assessment and is useful for protein structure prediction. The webserver and software packages of Qprob are available at: http://calla.rnet.missouri.edu/qprob/.

## Introduction

The number of protein sequences has grown exponentially during the last few decades because of the wide application of high-throughput next-generation sequencing technologies [1]. This ensures the importance of computational methods in bioinformatics and computational biology that are much cheaper and faster than experimental methods [2] for annotating the structure and function of these protein sequences [2-8]. A lot of progress has been made recently for protein structure prediction in terms of both template-based modeling and template-free modeling assisted by the Critical Assessment of Techniques for Protein Structure Prediction (CASP). During the prediction of protein structure, one important task is protein model quality assessment. The model quality assessment problem can be defined as ranking structure models of a protein without knowing its native structure, which is commonly used to rank and select many alternative decoy models generated by modern protein structure prediction methods. In general, there are two different kinds of protein quality assessment (QA) methods: single-model quality assessment [9-14] and consensus model quality assessment [15-17]. According to the previous CASP experiments, the consensus quality assessment methods usually perform better than the single-model quality assessment methods, especially when there is a good consensus in the model pool. However, it is also known that the consensus quality assessment may fail badly when there are a large portion of low-quality models in the model pool that are similar to each other [9]. Moreover, consensus quality assessment methods are very slow when there are more than tens of thousands of models to assess. Therefore, more and more single-model methods are developed to address these problems. Currently, most single-model methods use the evolution information [18], residue environment compatibility [19], structural features and physics-based knowledge [11-14,20-22] as features to assess model quality. Some hybrid methods also try to combine the single-model and consensus methods to achieve good performance [2,4]. In the work, we develop a single-model QA method (Qprob) that combines structural, physicochemical, and energy features extracted from a single model to improve protein model quality assessment. To the best of our knowledge, Qprob is the first method that estimates the errors of these features and integrates them by probability density functions to assess model quality assessment.

Specifically, we benchmarked four protein energy scores in combination with seven physicochemical and structural features. We quantify the effectiveness of these features on the PISCES[23] database, and calibrate their weights for combination. The probability density function of the errors between predicted quality scores and real GDT-TS scores for each feature is generated, assuming that the error between each predicted score and the real score roughly obeys the normal distribution. By combining the probability density distributions of all features, we can predict the global quality score of a model. The performance of Qprob is similar to the state-of-the-art single-model performance during the blind CASP11 experiment, which demonstrates the effectiveness of the probability density distribution based quality assessment methods.

The paper is organized as follows. In the Methods Section, we describe each feature and the calculation of the global quality assessment score in detail. In the Result Section, we report the performance of our method in the CASP11 experiment. In the Discussion Section, we summarize the results and conclude with the direction of future works.

## Results

### Feature normalization results

We use 11 feature scores in total in our method, and there is no need to do normalization for most of them. However, some features, especially the energy scores are dependent on the sequence length, and not in the range of 0 and 1. So we need to normalize these scores before using them. We use PISCES database to benchmark and normalize three scores (DFIRE2 score, RWplus score, and RF_CB_SRS_OD score) to remove length dependency. The sequence identity cutoff between any two sequences in the PISCES database is 20%, the resolution cutoff of structure is 1.8 Angstroms, and the R-factor cutoff of structure is 0.25. **Figure 1** shows the plot of the three original energy scores (DFIRE2, RWplus, and RF_CB_SRS_OD scores) versus protein length. We use linear regression to fit the energy score with the protein length (see the linear line in Figure 1). The following linear function formula describes the relationship between protein sequence length and the energy score:

$$\begin{cases} Dfire\ score & = -1.971 * L + 37.746 \\ RWplus\ score & = -232.6 * L + 6589.5 \\ RF\_CB\_SRS\_OD\ score & = -0.4823 * L + (-15.9066) \end{cases} \quad (1)$$

L is the protein sequence length. Based on these linear relationships, to normalize these scores into the range of 0 and 1, we use the following formula:

$$\begin{cases} Norm\_S_{Dfire} & = \frac{-P_{Dfire\ score}}{1.971*L} \\ Norm\_S_{RWplus} & = \frac{-P_{RWplus\ score}}{232.6*L} \\ Norm\_S_{RF\_CB\_SRS\_OD} & = \frac{700-P_{RF\_CB\_SRS\_OD\ score}}{1000+0.4823*L} \end{cases} \quad (2)$$

$P_{Dfire\ score}$ is the predicted DFIRE2 score, $P_{RWplus\ score}$ is the predicted RWplus score, and $P_{RF\_CB\_SRS\_OD\ score}$ is the predicted RF_CB_SRS_OD score. $P_{Dfire\ score}$ is set to the range of -1.971*L and 0, $P_{RWplus\ score}$ is set to the range of -232.6*L and 0, and $P_{RF\_CB\_SRS\_OD\ score}$ is set to the range of 0.4823*L-300 and 700 based on the benchmark of all scores in CASP9 targets so that most of models are in this range.

**Feature error estimation results**

We use each of 11 feature scores including three normalized energy scores to predict the quality score of the models of CASP9 targets, and the difference between predicted score and real GDT-TS score for each model is used estimate the probability of the error of each feature. **Figure 2** shows the probability density distribution of all 11 features, respectively. The x-axis is the error between predicted score and real GDT-TS score, and the y-axis is the probability density distribution of the error. The mean and standard deviation is also listed in the figures. We use a normal distribution to fit these errors. According to the results, ModelEvaluator score has the mean -0.0219, which is the closest to the real average GDT-TS score. In addition, it has the minimum standard deviation, which suggests it is the most stable feature for evaluating the global model quality. In contrast, the Euclidean compact score has the maximum absolute mean error (0.4119), showing it is most different from the real GDT-TS score.

**Global quality assessment results**

Qprob was blindly tested on CASP11 as MULTICOM-NOVEL server, and was used for the human tertiary structure predictor MULTICOM. MUTLCIOM is officially ranked 3rd out of 143 predictors according to the total scores of the first models predicted. According to the analysis result by removing each QA method from MULTICOM, the removal of Qprob causes the biggest decrease in the average Z-score of top one models selected by MULTICOM method (Z-score from 1.364 to 1.321) [2,4], showing Qprob makes big contribution to MUTLCOM. Our method is one of the best single-model QA method based on the CASP official evaluation [24] and our own evaluations reported in **Table1** and **Table2**.

**Table 1** depicts the per-target average correlation, average GDT-TS loss, average spearman's correlation, and average kendall tau correlation of our method Qprob and other pure single-model QA methods on Stage 1 (sel20) CASP11 datasets. These scores are calculated by comparing the model quality scores predicted by each of these methods with the real model quality scores. We also report the p-value of the pairwise Wilcoxon signed ranked sum test for the difference of loss/correlation between Qprob and other pure single-model QA methods. The method in the table are ordered by the average GDT-TS loss (the difference of GDT-TS score of the best model and predicted top 1 model) that assess a method's capability of selecting good models. According to **Table 1**, Qprob is ranked third based on the average GDT-TS loss on Stage 1 CASP11 datasets. According to 0.01 significant threshold of p-value, there is no significant difference between Qprob and the two state-of-the-art QA methods ProQ2 and ProQ2-refine in terms of both correlation and loss. The difference on average Spearman's correlation and Kendall tau correlation is also small between Qprob and the other two top performing methods. Other than CASP11 QA server predictors, we also compare Qprob with five single-model QA scores that are highlighted in bold in **Table 1**. The five scores are ModelEvaluator score, Dope score, DFIRE2 score, RWplus score, and RF_CB_SRS_OD score. The result shows Qprob performs better than these scores in terms of both correlation and loss. Moreover, the difference of correlation between Qprob and four QA scores (Dope score, DFIRE2 score, RWplus score, and RF_CB_SRS_OD score) is significant, and the difference of loss between Qprob and three QA scores (DFIRE2 score, RWplus score, and RF_CB_SRS_OD score) is significant according to 0.01 significance threshold. Finally, we also calculate the performance of the baseline consensus QA method DAVIS_consensus whose correlation and loss is 0.798 and 0.052 respectively. Not surprisingly, the performance of our single-model method Qprob is worse than DAVIS_consensus method, and the difference is significant. The p-value of difference in correlation and loss is 1.4e-12 and 1.6e-4 respectively. The difference is more significant between Qprob and the start-of-the-art consensus QA method Pcons-net[25] whose correlation and loss is 0.811 and 0.024, with p-value 1.93e-14 and 1.61e-6 respectively. We also evaluate the performance of Qprob and other QA methods on Stage2 (top150) CASP11 datasets in **Table 2**. Each target in the Stage2 CASP11 data has about 150 models. Qprob ranked second among all pure single-model QA methods based on the average loss metric. The difference between Qprob and ProQ2 is not significant in terms of both correlation and loss, i.e., p-value is 0.2387 and 0.8636 respectively, showing Qprob achieved close to state-of-the-art model selection ability. Comparing Qprob with five scores (ModelEvaluator score, Dope score, DFIRE2 score, RWplus score, and RF_CB_SRS_OD score), the difference of correlation between Qprob and four scores (ModelEvaluator score, Dope score, DFIRE2 score, and RWplus score) is significant, and the difference of loss between Qprob and two scores (DFIRE2 score, and RF_CB_SRS_OD score) is significant according to p-value threshold 0.01. In addition, we also compare the performance of Qprob with baseline consensus method DAVIS_consensus on Stage2 (top150) CASP11 datasets. The per-target average correlation of DAVIS_consensus is 0.57, which is better than Qprob (with correlation 0.381). The difference of correlation is significant according to p-value 6.14e-4. However, the per-target loss of Qprob (i.e. 0.068) is better than DAVIS_consensus (i.e. 0.073). Although the difference of loss is not very significant (p-value 0.11), this shows Qprob performs at least as well as the consensus method on Stage2 CASP11 datasets. Moreover, compared with the top performing consensus QA method Pcons-net loss (i.e. 0.049), the difference of loss between Qprob and Pcons-net is still not very significant (p-value 0.19). To illustrate the model selection ability of the QA methods on hard targets whose model pool contains mostly low quality models, we evaluate the performance of Qprob and several top performing single-model/consensus QA methods on the template free

CASP11 targets. We calculate the summation of Z-score for the top 1 model selected by each QA method. The result is shown in **Figure 3**. **Figure 3A** shows the performance of each method on Stage1 of CASP11 datasets. The single-model QA methods are in bold. The consensus QA methods have relatively better performance, where, the baseline pairwise method DAVIS_consensus gets the highest Z-score. **Figure 3B** shows the performance of each QA method on Stage 2 of CASP11 datasets. It is very interesting to see that the single-model QA methods have relatively better performance than consensus QA methods. Especially, our method Qprob and VoromQA have the highest Z-score comparing with other single QA methods. Another interesting finding is the pairwise method DAVIS_consensus has Z-score around 0, suggesting its performance is close to a random predictor. These results demonstrate the value of single-model QA method in selecting models of hard targets. The hybrid method MULTICOM-CONSTRUCT that combines both single-model and consensus methods ranks third, showing the combination of the two kinds of methods is also quite useful for model selection.

**Discussion**

In this paper, we introduce a novel single-model QA method Qprob. Different from other single-model QA methods, Qprob estimates the prediction error estimation of several different physicochemical, structural and energy feature scores, and use the combination of probability density distribution of the errors for the global quality assessment. We blindly tested our method in the CASP11 experiment, and it was ranked as one of the best single-model QA method based on the CASP official evaluation and our own evaluations. In particular, the good performance of our method on template free targets demonstrates its good capability of selecting models for hard targets. Furthermore, the method made valuable contribution to our MULTICOM human tertiary structure predictor - one of the best human predictors among all server and human predictors in CASP11. These results demonstrate the broad application of our method in model selection and protein structure prediction.

**Methods**

In this section, we describe the calculation of 11 features, how to generate the probability density distributions of the prediction errors of these features, and how to combine these features for protein model quality assessment.

**Feature generation**

Our method uses structural / sequence features extracted from a structural model and its protein sequence, physicochemical characteristic of the structural model [21], and four energy scores for predicting the global quality score of the model. The features include:

1. The RF_CB_SRS_OD score [11] is an energy score for evaluating the protein structure based on statistical distance dependent pairwise potentials. The score is normalized into the range of 0 and 1(see the normalization protocol in the Result section).

2. The secondary structure similarity score is calculated by comparing the secondary structure predicted by Spine X [26] from a protein sequence and those of a model parsed by DSSP [27].

3. The secondary structure penalty percentage score is calculated by the following formula:

$$S_{penalty} = \frac{F_H + F_S}{N} \quad (3)$$

$F_H$ is the total number of predicted helix residues matching with the one parsed by DSSP. $F_S$ is the total number of predicted beta-sheet residues matching with the ones parsed by DSSP. $N$ is the sequence length.

4. The Euclidean compact score is used to describe the compactness of a protein model. It is calculated by the following formula:

$$S_{Eucli} = \frac{\sum Eucli(i,j)}{\sum 3.8 * |i-j|} \quad (4)$$

i and j is the index of any two amino acids, and Eucli(i,j) is the Euclidean distance of amino acid i and j in the structural model.

5. The surface score for exposed nonpolar residues describes the percentage of exposed area of the nonpolar residues, and is calculated as follows:

$$S_{surf} = \frac{\sum SE_i}{\sum S_i} \tag{5}$$

$S_i$ is the exposed area of residue i parsed by DSSP, and $SE_i$ is the exposed area of nonpolar residue i. The $SE_i$ is set to 0 if residue i is polar.

6. The exposed mass score describes the percentage of mass of exposed residues, and is calculated as follows:

$$S_{mass} = \frac{\sum STN_i * M_i}{\sum S_i * M_i} \tag{6}$$

$S_i$ is the exposed area of residue i parsed by DSSP, $STN_i$ is the total area of nonpolar residue i, and $M_i$ is the total mass of residue i.

7. The exposed surface score describes the percentage of area of the residues exposed, and is calculated as follows:

$$S_{exposed\ surface} = \frac{\sum S_i}{\sum ST_i} \tag{7}$$

$ST_i$ is the total area of residue i parsed by DSSP, and $S_i$ is the exposed area of residue i.

8. The solvent accessibility similarity score is calculated by comparing solvent accessibility predicted SSpro4[28] from the protein sequence and those of a model parsed by DSSP [27].

9. The RWplus score [12] is an energy score evaluating protein models based on distance-dependent atomic potential. The score is normalized to the range of 0 and 1.

10. The ModelEvaluator score [13] is a score evaluating protein models based on structural features and support vector machines.

11. The Dope score [14] is an energy score evaluating protein models based on the reference state of non-interacting atoms in homogeneous sphere. The score is normalized to the range of 0 and 1.

**Feature errors estimation**

We calculate all feature scores for the models of 99 CASP9 targets, which in total have 22016 models. The feature error is calculated for each model using the following formula:

$$FE_{i,j} = F_{i,j} - R_j \tag{8}$$

$FE_{i,j}$ is the error estimate of feature i on model j, $F_{i,j}$ is the predicted score of feature i on model j, and $R_j$ is the real GDT-TS score of model j. Based on these errors, we calculate the mean $M_i$ and standard deviation $SD_i$ for each feature i as follows:

$$\begin{cases} M_i = \frac{\sum_{j=1}^{N} FE_{i,j}}{N} \\ SD_i = \sqrt{\frac{\sum_{j=1}^{N}(FE_{i,j}-M_i)^2}{N}} \end{cases} \tag{9}$$

i is in the range of 1 and 11 which represent all 11 features. N is the total number of models. The feature error estimation statistics (mean and standard deviation of each feature) is used for global model quality score assessment.

**Feature weight estimation**

We use a method similar to EM algorithm for estimating the weight of each feature for combination. The weight for each feature is used to adjust the feature distributions. The algorithm has three steps as following:

1. Initialization: Randomly assign a weight to each feature. The weight value is chosen from the range -0.8 to 0.8 with the step size of 0.01, and assign the minimum average GDT-TS loss (Min-Loss) to 1.

2. Expectation step: calculating the per-target average loss using the current weight value set W benchmarked on CASP9 targets.

3. Maximization step: trying different weight values for feature i while fixing the weight of all other features. For each weight w for feature i, get the average GDT-TS loss from step 2, and updating the Min-Loss (the minimum loss) if it is less than the current value of Min-Loss. The current weight value set W is updated if a new weight w of features i that lowers the loss is found. Repeat step 3 for the next feature i+1 until all the features are used. And repeat the process until the current weight value set W is not changed.

After applying this algorithm on the CASP9 dataset, we obtain a weight value set W which has the minimum average GDT-TS loss. The best weights for 11 features are:

[0.03,0.09,0.04,0.08,0.08,0.01,0.03,0.10,0.00,0.09,-0.02].

**Model quality assessment based on probability density function**

Given a protein model, we first calculate feature score $Pre_i$ for each feature i (i is in the range [1 and 11]). And then we calculate the adjusted score (an estimation of the real score) by $Adjust\_pre_i = Pre_i - M_i$, while the mean $M_i$ and standard deviation $SD_i$ of each feature i has been calculated in the feature errors estimation as described above. We use the following probability density function of global quality $X_i$ for each feature i (the mean is $Adjust\_pre_i$ and standard deviation is $SD_i$) to quantify the predicted global quality score $X_i$:

$$P_i(X_i) = \frac{e^{-\frac{(X_i - Adjust\_pre_i)^2}{2*SD_i^2}}}{\sqrt{2\pi} SD_i} \qquad (10)$$

We normalize the probability score to convert it into the range of 0 and 1 with the following formula:

$$P\_norm_i(X_i) = \frac{P_i(X_i)}{P_i(Adjust\_pre_i)} \qquad (11)$$

The final global quality score is calculated by combining all 11 normal distributions from each feature prediction. Given a value X in the range of 0 and 1, we calculate the combined probability score as follows:

$$P\_combine(X) = \sum_{i=1}^{i=11}(P_{norm_i}(X + W_i)) \qquad (12)$$

The value X that has the maximum combined probability score $P\_combine(X)$ is assigned as the global quality score for the model. Here, $W_i$ is the weight of feature i.

**References**

1   Li, J., Cao, R. & Cheng, J. A large-scale conformation sampling and evaluation server for protein tertiary structure prediction and its assessment in CASP11. BMC bioinformatics **16**, 337 (2015).
2   Cao, R., Bhattacharya, D., Adhikari, B., Li, J. & Cheng, J. Large-scale model quality assessment for improving protein tertiary structure prediction. Bioinformatics **31**, i116-i123 (2015).
3   Wang, Z., Cao, R. & Cheng, J. Three-level prediction of protein function by combining profile-sequence search, profile-profile search, and domain co-occurrence networks. BMC bioinformatics **14**, S3 (2013).
4   Cao, R., Bhattacharya, D., Adhikari, B., Li, J. & Cheng, J. Massive integration of diverse protein quality assessment methods to improve template based modeling in CASP11. Proteins: Structure, Function, and Bioinformatics, doi:10.1002/prot.24924 (2015).
5   Cao, R. & Cheng, J. Integrated protein function prediction by mining function associations, sequences, and protein-protein and gene-gene interaction networks. Methods **93**, 84-91 (2016).
6   Cao, R. & Cheng, J. Deciphering the association between gene function and spatial gene-gene interactions in 3D human genome conformation. BMC genomics **16**, 880 (2015).
7   Adhikari, B., Bhattacharya, D., Cao, R. & Cheng, J. CONFOLD: Residue‐residue contact‐guided ab initio protein folding. Proteins: Structure, Function, and Bioinformatics **83**, 1436-1449 (2015).
8   Li, J. et al. The MULTICOM protein tertiary structure prediction system. Protein Structure Prediction **1137**, 29-41, doi:10.1007/978-1-4939-0366-5_3 (2014).
9   Cao, R., Wang, Z. & Cheng, J. Designing and evaluating the MULTICOM protein local and global model quality prediction methods in the CASP10 experiment. BMC structural biology **14**, 13 (2014).
10  Cao, R., Wang, Z., Wang, Y. & Cheng, J. SMOQ: a tool for predicting the absolute residue-specific quality of a single protein model with support vector machines. BMC bioinformatics **15**, 120 (2014).

11   Rykunov, D. & Fiser, A. Effects of amino acid composition, finite size of proteins, and sparse statistics on distance‐dependent statistical pair potentials. *Proteins: Structure, Function, and Bioinformatics* **67**, 559-568 (2007).

12   Zhang, J. & Zhang, Y. A novel side-chain orientation dependent potential derived from random-walk reference state for protein fold selection and structure prediction. *PLoS One* **5**, e15386, doi:10.1371 (2010).

13   Wang, Z., Tegge, A. N. & Cheng, J. Evaluating the absolute quality of a single protein model using structural features and support vector machines. *Proteins* **75**, 638-647, doi:10.1002/prot.22275 (2009).

14   Shen, M. y. & Sali, A. Statistical potential for assessment and prediction of protein structures. *Protein Science* **15**, 2507-2524 (2006).

15   McGuffin, L. The ModFOLD server for the quality assessment of protein structural models. *Bioinformatics* **24**, 586 - 587 (2008).

16   Wang, Q., Vantasin, K., Xu, D. & Shang, Y. MUFOLD-WQA: a new selective consensus method for quality assessment in protein structure prediction. *Proteins* **79**, 185 - 195 (2011).

17   McGuffin, L. & Roche, D. Rapid model quality assessment for protein structure predictions using the comparison of multiple models without structural alignments. *Bioinformatics* **26**, 182 - 188 (2010).

18   Kalman, M. & Ben-Tal, N. Quality assessment of protein model-structures using evolutionary conservation. *Bioinformatics* **26**, 1299 - 1307 (2010).

19   Liithy, R., Bowie, J. & Eisenberg, D. Assessment of protein models with three-dimensional profiles. *Nature* **356**, 83 - 85 (1992).

20   Ray, A., Lindahl, E. & Wallner, B. Improved model quality assessment using ProQ2. *BMC bioinformatics* **13**, 224 (2012).

21   Mishra, A., Rao, S., Mittal, A. & Jayaram, B. Capturing native/native like structures with a physico-chemical metric (pcSM) in protein folding. *Biochimica et Biophysica Acta (BBA)-Proteins and Proteomics* **1834**, 1520-1531 (2013).

22   Benkert, P., Biasini, M. & Schwede, T. Toward the estimation of the absolute quality of individual protein structure models. *Bioinformatics* **27**, 343-350 (2011).

23   Wang, G. & Dunbrack, R. L. PISCES: a protein sequence culling server. *Bioinformatics* **19**, 1589-1591 (2003).

24   Kryshtafovych, A. et al. Methods of model accuracy estimation can help selecting the best models from decoy sets: assessment of model accuracy estimations in CASP11. *Proteins: Structure, Function, and Bioinformatics*, doi:10.1002/prot.24919 (2015).

25   Wallner, B. & Elofsson, A. Identification of correct regions in protein models using structural, alignment, and consensus information. *Protein Sci* **15**, 900 - 913 (2009).

26   Faraggi, E., Zhang, T., Yang, Y., Kurgan, L. & Zhou, Y. SPINE X: improving protein secondary structure prediction by multistep learning coupled with prediction of solvent accessible surface area and backbone torsion angles. *Journal of computational chemistry* **33**, 259-267 (2012).

27   Kabsch, W. & Sander, C. Dictionary of protein secondary structure: pattern recognition of hydrogen‐bonded and geometrical features. *Biopolymers* **22**, 2577-2637 (1983).

28   Cheng, J., Randall, A. Z., Sweredoski, M. J. & Baldi, P. SCRATCH: a protein structure and structural feature prediction server. *Nucleic Acids Research* **33**, W72-W76 (2005).

## Acknowledgments

We thank Audrey Van Leunen for checking the English grammars, and to the CASP11 organizers and assessors for providing comprehensive data. The work is supported by US National Institutes of Health (NIH) grant (R01GM093123) to JC.

## Author contributions statement

RC and JC conceived and designed the method and the system. RC implemented the method, built the system, and carried out the CASP experiments. RC and JC wrote the manuscript. All the authors approved the manuscript.

## Additional Information

**Competing financial interests**

The authors declare no competing financial interests.

## Figure legends

**Figure 1. The relationship of three energy scores (DFIRE2, RWplus, and RF_CB_SRS_OD scores) and sequence length on PISCES database.**

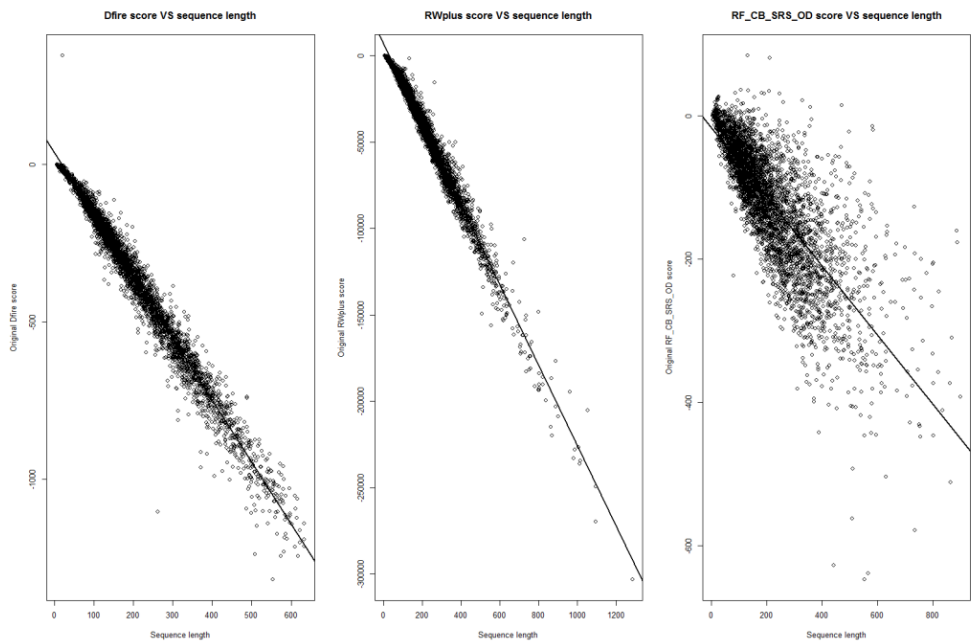

**Figure 2. The probability density distributions for the error estimation of all 11 feature scores.**

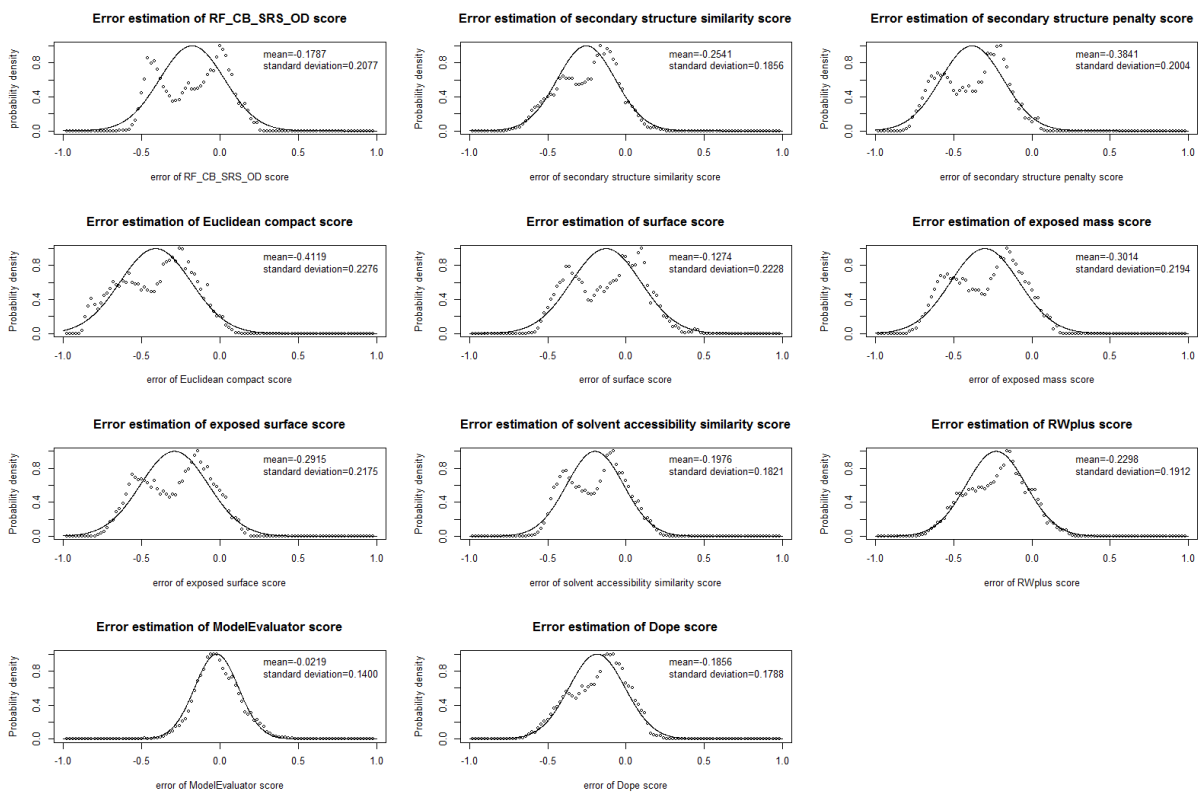

**Figure 3. The summation of Z-score for the top 1 models selected by each method.**

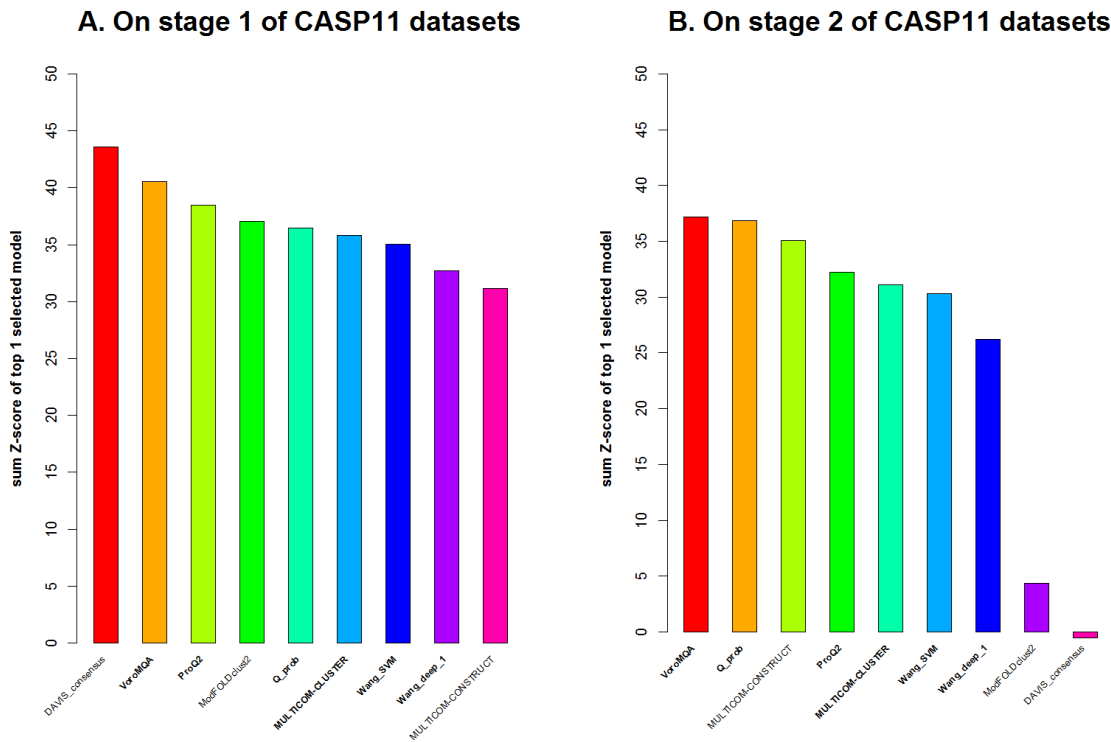

**Table 1. The per-target average correlation, average loss, average Spearman's correlation, average Kendall tau score, and total number of evaluated targets of Qprob and other pure single-model QA methods on sel20 CASP11 dataset. The p-value of pairwise Wilcoxon signed ranked sum test for the difference of loss and correlation of Qprob against other methods is listed for comparison. Five single-model QA methods which did not attend CASP11 are also listed and highlighted in bold**.

| QA Methods | Ave. corr. | Ave. loss | Ave. spearman | Ave. kendall. | p-value loss | p-value corr. | # |
|---|---|---|---|---|---|---|---|
| ProQ2 | 0.643 | 0.090 | 0.506 | 0.379 | 0.9776 | 0.2755 | 84 |
| ProQ2-refine | 0.653 | 0.093 | 0.535 | 0.402 | 0.9935 | 0.01756 | 84 |
| Qprob | 0.631 | 0.097 | 0.517 | 0.389 | - | - | 84 |
| **ModelEvaluator** | 0.6 | 0.097 | 0.47 | 0.353 | 0.9224 | 0.2678 | 84 |
| VoroMQA | 0.561 | 0.108 | 0.426 | 0.318 | 0.288 | 8.61E-05 | 84 |
| Wang_SVM | 0.655 | 0.109 | 0.535 | 0.401 | 0.09109 | 0.003131 | 84 |
| **Dope** | 0.542 | 0.111 | 0.416 | 0.316 | 0.06388 | 9.56E-10 | 84 |
| Wang_deep_2 | 0.633 | 0.115 | 0.514 | 0.388 | 0.03468 | 0.2755 | 84 |
| Wang_deep_3 | 0.626 | 0.117 | 0.513 | 0.388 | 0.008288 | 0.6034 | 84 |
| Wang_deep_1 | 0.613 | 0.128 | 0.517 | 0.386 | 0.000559 | 0.403 | 84 |
| **DFIRE2** | 0.502 | 0.135 | 0.388 | 0.284 | 0.000589 | 1.08E-12 | 84 |
| **RWplus** | 0.536 | 0.135 | 0.433 | 0.323 | 0.002436 | 6.52E-11 | 84 |
| FUSION | 0.095 | 0.154 | 0.133 | 0.099 | 0.001565 | 4.05E-13 | 84 |
| raghavagps-qaspro | 0.35 | 0.156 | 0.263 | 0.187 | 0.00019 | 6.02E-12 | 84 |
| **RF_CB_SRS_OD** | 0.486 | 0.162 | 0.357 | 0.256 | 0.000114 | 4.56E-09 | 84 |

**Table 2. The per-target average correlation, average loss, average Spearman's correlation, average Kendall tau score, and total number of evaluated targets of Qprob and several other pure single-model QA methods on Stage2 CASP11 dataset. The p-value of pairwise Wilcoxon signed ranked sum test for the difference of loss and correlation of Qprob against other methods is listed for comparison. Five single-model QA methods which did not attend CASP11 are also listed and highlighted in bold**.

| QA Method | Ave. corr. | Ave. loss | Ave. spearman | Ave. kendall. | p-value loss | p-value corr. | # |
|---|---|---|---|---|---|---|---|
| ProQ2 | 0.372 | 0.058 | 0.366 | 0.256 | 0.2387 | 0.8636 | 83 |
| Qprob | 0.381 | 0.068 | 0.387 | 0.272 | - | - | 83 |
| VoroMQA | 0.401 | 0.069 | 0.386 | 0.269 | 0.4335 | 0.5864 | 83 |
| ProQ2-refine | 0.37 | 0.069 | 0.375 | 0.264 | 0.2442 | 0.9656 | 83 |
| **ModelEvaluator** | 0.324 | 0.072 | 0.305 | 0.212 | 0.002554 | 0.3084 | 83 |
| **Dope** | 0.304 | 0.077 | 0.324 | 0.228 | 1.59E-07 | 0.74 | 83 |
| **RWplus** | 0.295 | 0.084 | 0.314 | 0.22 | 7.00E-09 | 0.11 | 83 |
| Wang_SVM | 0.362 | 0.085 | 0.351 | 0.245 | 0.4774 | 0.1502 | 83 |
| raghavagps-qaspro | 0.222 | 0.085 | 0.205 | 0.139 | 3.07E-07 | 0.006219 | 83 |
| Wang_deep_2 | 0.307 | 0.086 | 0.298 | 0.208 | 0.000593 | 0.03628 | 83 |
| Wang_deep_1 | 0.302 | 0.089 | 0.293 | 0.203 | 0.000911 | 0.04544 | 83 |
| **DFIRE2** | 0.235 | 0.091 | 0.253 | 0.175 | 6.15E-11 | 0.004036 | 83 |
| Wang_deep_3 | 0.302 | 0.092 | 0.29 | 0.202 | 0.000469 | 0.008166 | 83 |
| **RF_CB_SRS_OD** | 0.36 | 0.097 | 0.35 | 0.243 | 0.06173 | 0.002035 | 83 |
| FUSION | 0.05 | 0.111 | 0.082 | 0.054 | 7.16E-11 | 5.82E-07 | 83 |